\documentclass[sigconf,nonacm]{acmart} 

\setcopyright{none}

\usepackage{subcaption}
\begin{document}
	
	\sloppy

\title[Capture the Flag for Team Construction in Cybersecurity] {Capture the Flag for Team Construction in Cybersecurity} 

\author{Sang-Yoon Chang$^*$\hspace{2.0em} Kay Yoon$^\mathsection$\hspace{2.0em} Simeon Wuthier$^*$ \hspace{2.0em} Kelei Zhang$^*$}
\affiliation{%
 \institution{University of Colorado Colorado Springs$^{*\mathsection}$}
 \institution{Computer Science Department$^*$ \hspace{3.0em} Department of Communication$^\mathsection$} 
  \institution{ \{schang2,\hspace{0.6em} kyoon,\hspace{0.6em} swuthier,\hspace{0.6em} jzhang5\}@uccs.edu }
  \country{Colorado Springs, CO, USA}
}
\renewcommand{\shortauthors}{S.Y. Chang, et al.} 

\begin{abstract}
  Team collaboration among individuals with diverse sets of expertise and skills is essential for solving complex problems. As part of an interdisciplinary effort, we studied the effects of Capture the Flag (CTF) game, a popular and engaging education/training tool in cybersecurity and engineering, in enhancing team construction and collaboration. 
  We developed a framework to incorporate CTF as part of a computer-human process for expertise recognition and role assignment and evaluated and tested its effectiveness through a study with cybersecurity students enrolled in a Virtual Teams course. In our computer-human process framework, the post-CTF algorithm using the CTF outcomes assembles the team (assigning individuals to teams) and provides the initial role assignments, which then gets updated by human-based team discussions. This paper shares our insights, design choices/rationales, and  analyses of our CTF-incorporated computer-human process framework. The students' evaluations revealed that the computer-human process framework was helpful in learning about their team members' backgrounds and expertise and assigning roles accordingly 
  made a positive impact on the learning outcomes for the team collaboration skills in the course. This experience report showcases the utility of CTF as a tool for expertise recognition and role assignments in teams and highlights the complementary roles of CTF-based and discussion-based processes for an effective team collaboration among engineering students.
\end{abstract}



\maketitle

\section{Introduction}

Team collaboration skills are becoming one of the most desired competencies in the contemporary workforce and considered fundamental to engineering professions \cite{Cabrera2017}. Team collaboration skills are particularly relevant for cybersecurity professionals, whose work requires coordination of collective efforts involving individuals with differing sets of knowledge, skills, and abilities when developing solutions for complex cybersecurity threats \cite{YoonChang2021}. At the core of effective team collaboration is team members' abilities to recognize each other's strengths and fully utilize them by delegating responsibilities to each other in line with each member's areas of expertise \cite{Hollingshead2011}.

This experience report describes our study of the impact of a computer-human process on enhancing team collaboration. Specifically, we developed a two-pronged team-knowledge sharing process designed to help teams configure their members' functional roles in a way that maximizes the utilization of individual members' knowledge. The first step of the process was a Capture the Flag (CTF) exercise in which team members solved challenges in disparate functional areas in the field of cybersecurity and demonstrated their competency in each area. Our framework then used the individual CTF performance data to optimize team assembly and individual role assignment. The second step of the process involved teams' collective discussion in which team members further shared their professional and academic experiences/expertise beyond what they demonstrated in the CTF exercise and adjusted the initial algorithmic role assignment. This two-pronged process for role assignment illustrates a unique approach integrating both algorithm-based and human-based mechanisms for improving team construction including team assembly and role assignments.

\section{Related Work}

\begin{figure*}[th]
\begin{subfigure}{0.48\textwidth}
\begin{center}
\includegraphics[width=0.92\columnwidth]{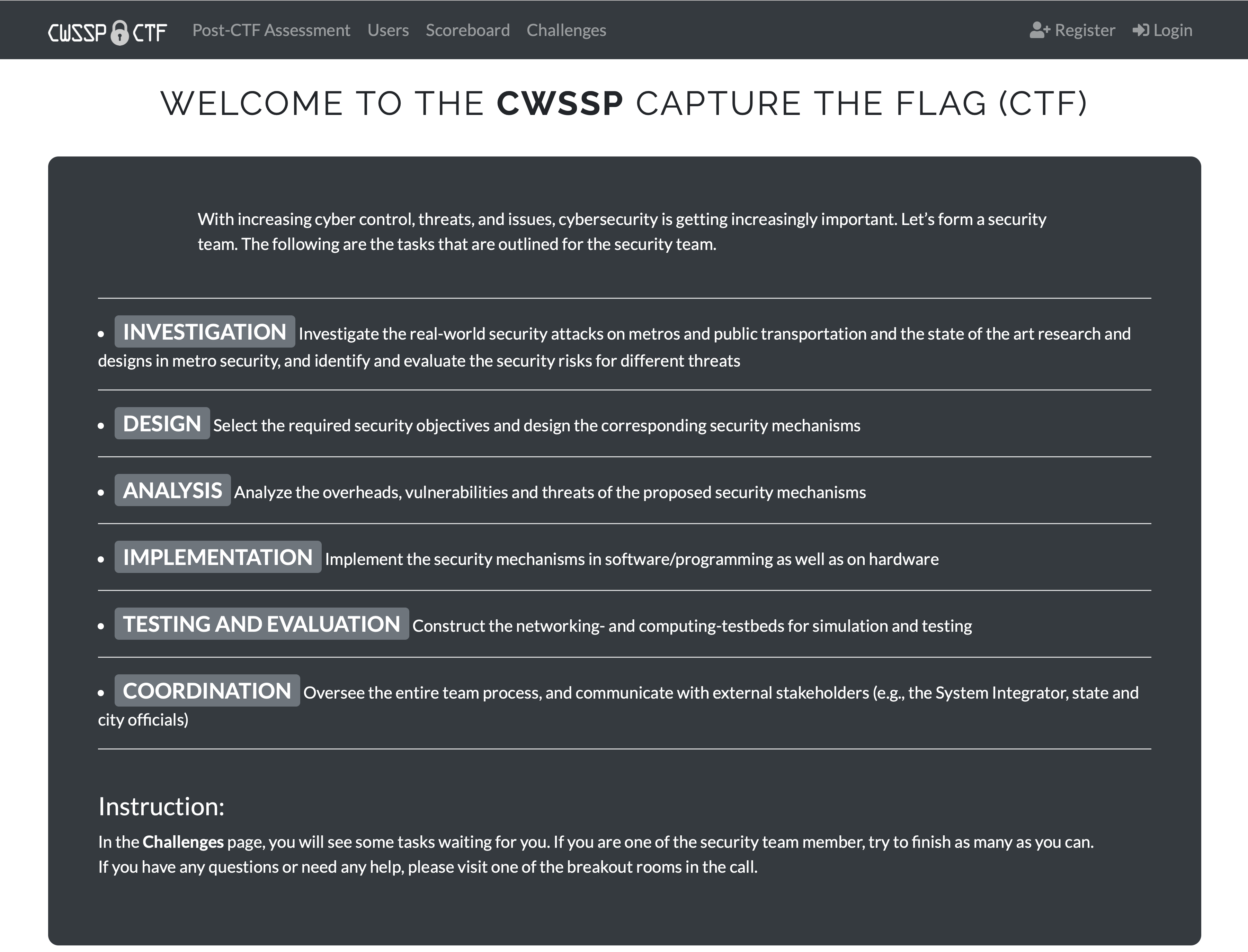} 
\caption{CTF role categories as shown to the participants} 
\label{fig:ctf_roles}
\end{center}
\end{subfigure}
\hfill
\begin{subfigure}{0.48\textwidth}
\begin{center}
\includegraphics[width=0.96\columnwidth]{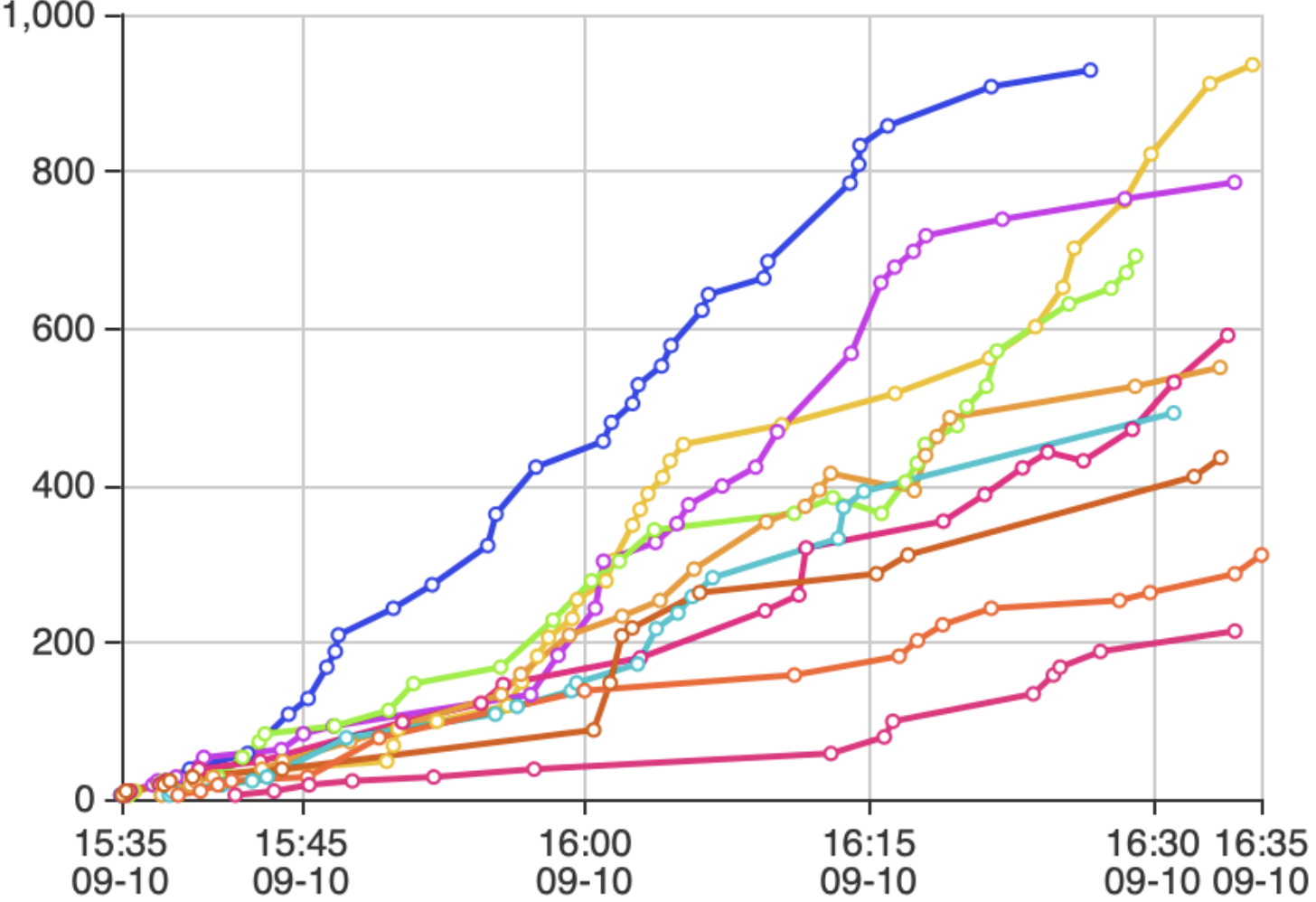} 
\caption{The participant scores summed across all roles/skills (in different curves) vs. time as progressed in the CTF game}
\hfill
\label{fig:ctf_scores}
\end{center}
\end{subfigure}
\hfill
\vspace{-0.1in}
\caption{CTF game, as shown to the participants during the game}
\label{fig:ctf_game}
\end{figure*}

\subsection{Team Collaboration and Role Assignment}

Research on team collaboration integrates a variety of disciplinary perspectives including psychology, communication, and management. One of the dominant streams of research on team collaboration is concerned with knowledge sharing processes and offers robust theoretical and empirical grounds showing that knowledge sharing processes are a key determinant for effective team collaboration. A theoretical lens we ground our current research on is Transactive Memory Systems (TMS) theory, which explains that groups develop a shared memory among its members regarding who is knowledgeable about what and use the memory system for delegating responsibilities to individual members \cite{Wegner1987}. Communication and interaction among group members serve as a critical vehicle through which group members learn about each other's expertise and develop their collective memory system. The more accurately group members recognize each other's knowledge base, the more likely they align each member's expertise with his/her roles and responsibilities and share knowledge more effectively \cite{Brandon2004, Hollingshead2011}. Empirical evidence has shown a robust positive relationship between TMS and team performance in a variety of task environments \cite{Ren2011}.    

An accurate recognition of team members' expertise requires accumulated shared experiences through which one's knowledge is demonstrated, performed, or evaluated over time. However, when a team is newly formed, its members rely on initial perceptions and judgements based on limited interactions to gauge each other's expertise, which often leads to stereotyping or inaccurate assessment of one's abilities \cite{Yoon2010}. Also, individuals can strategically reveal, hide, or exaggerate one's expertise to meet a variety of individual goals \cite{Yoon2019}. Therefore, it is challenging to ensure accurate recognition of expertise in newly formed teams due to the psychological and social processes that are prone to errors. Although recent empirical studies identify the ways to enhance the accuracy of expertise recognition in teams \cite{Yoon2021}, it is still limited in offering tangible interventions for teams to implement. To address this gap, our current research developed a CTF exercise for teams that could serve as expertise cues based on individual performance data and used an algorithmic mechanism to assemble teams and assign roles for them. We argue that, in combination with team's verbal discussion, a performance-based algorithm enhances the accuracy of expertise recognition and the alignment between roles and expertise.

\subsection{CTF in Education}

CTF is popularly used in cybersecurity and engineering, 
including in classrooms and labs~\cite{beltran_experiences_2018,prabawa_using_2017}, 
for training security professionals~\cite{cherinka_using_2019}, for pedagogical research based on live CTF competition~\cite{katsantonis_conceptual_2017}, and for taxonomy framework research for training and education~\cite{knupfer_cyber_2020}. 
While existing practices incorporate CTF for talent acquisition and job placements, e.g., National Cyber League and MITRE~\cite{cherinka_using_2019}, the analyses of the impact of these methods are lacking, which motivated our work. 
In particular, we measure the impact of the CTF exercise on knowledge sharing in teams and examine how it serves as a tool for facilitating expertise recognition.

\section{Background and Context}


Our study was conducted in a three-week long online course "Virtual Teams", designed for cybersecurity students (pursuing degrees in cybersecurity in the engineering college) enrolled in a NSF-funded scholarship program\footnote{Colorado-Washington Security Scholars Program (CWSSP) is a NSF-funded CyberCorps Scholarship for Service (SFS) program for cybersecurity students at bachelor's, Master's, and PhD levels hosted by University of Colorado Colorado Springs and University of Washington Tacoma~\cite{bai2021}.}. 
The scholarship program emphasizes the importance of teamwork skills for future cybersecurity professionals and requires all student scholars enrolled in the program to take the Virtual Teams course. One of the assignments in the course requires students to coordinate and develop areas of responsibilities as a team in a hypothetical situation where a team of cybersecurity professionals needs to develop a security measure for a local mass transit system. This team assignment was designed to offer students an opportunity to apply the theoretical concepts related to team knowledge sharing and expertise recognition to a real-life situation and to experience how best to recognize team members' expertise and assign roles to each member as a team.

The team assignment required each student team to share each other's academic, professional backgrounds, research abilities, software and hardware skills, communication skills, and other expertise and using it to assign a functional role to each person. Before integrating the CTF game exercise in our study in the past offerings of the Virtual Teams course, the student teams relied solely on their team discussion when learning about each other's expertise and decided on who would be most capable of performing which functional role. By adding the CTF exercise, student teams were offered an additional data point regarding their team members' abilities through the CTF exercise and had an opportunity to combine the CTF performance data with their team's discussion to determine each member's functional roles. To test the effect of the CTF exercise on their team knowledge sharing and role assignment, we compared student teams with and without the CTF exercise in terms of their response to a questionnaire that measured their team collaboration efficacy.

\noindent \textbf{Participants} \hspace{0.15in} 
The participants were university students in the engineering college pursuing security degrees. In 2021, ten students (nine males and one female; one Ph.D. student, four Master's, and five undergraduates) participated in the Virtual Teams course and the CTF exercise. In 2020 where the CTF exercise was not implemented, seven students (six males and one female; one Ph.D. student, two Master's, and four undergraduates) participated in the Virtual Teams course. 

Because our CTF's main purpose is for education and teaching, our research did not require approval from the Institutional Review Board (IRB).

\begin{figure*}[th]
	\includegraphics[width=0.72\textwidth]{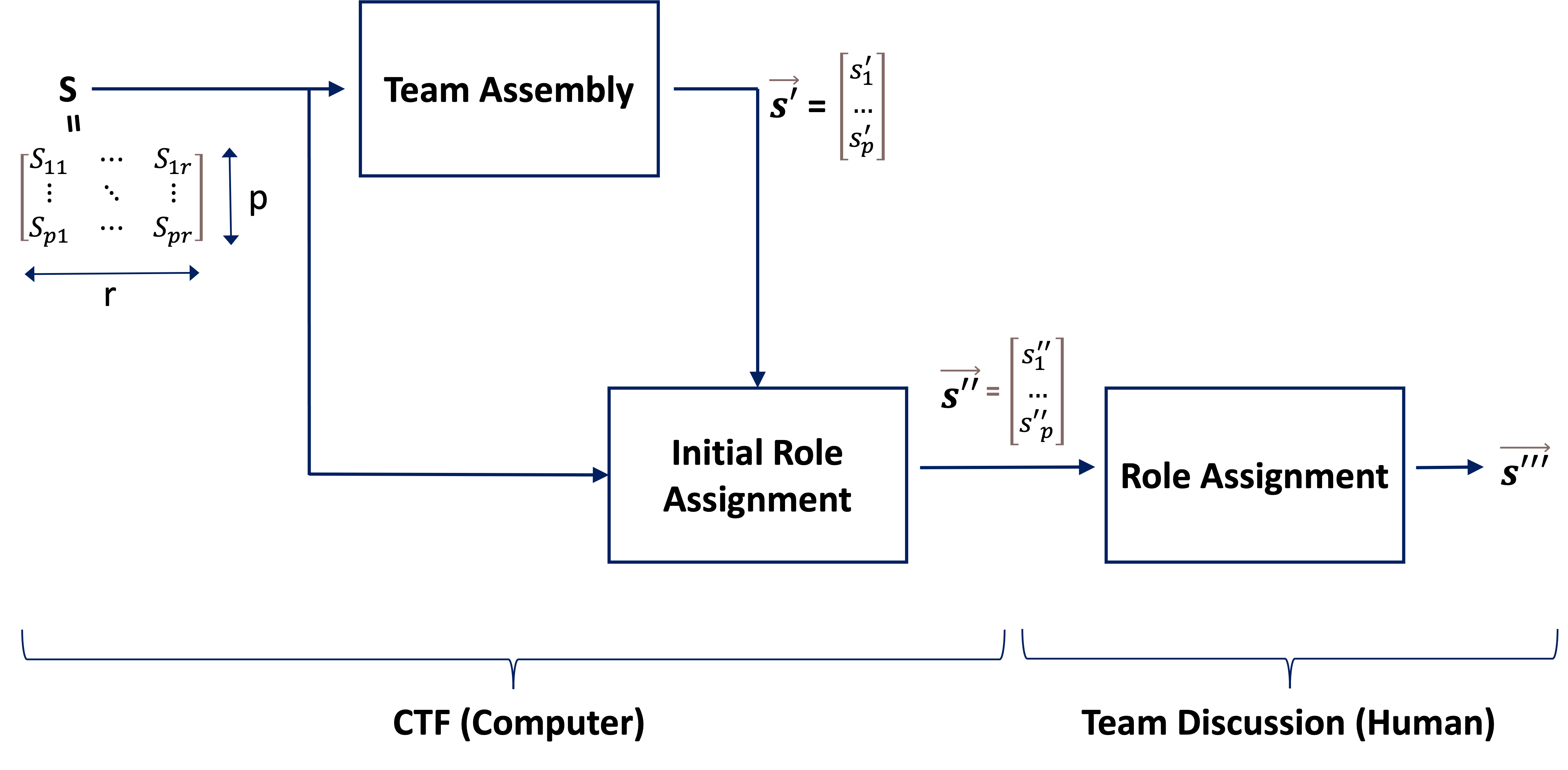}
	\caption{Overview including the CTF and team discussion for team assembly (top row) and the role assignment (bottom row). The CTF score matrix $S$ is used for team assembly, resulting in $\vec{s'}$ including the team information as its elements. The initial role assignment using the CTF scores ($S$) and the team assembly ($\vec{s'}$) results in $\vec{s''}$, followed by the team discussion to update the role assignment in $\vec{s'''}$; both $\vec{s''}$ and $\vec{s'''}$ include the role information as their elements.}
	\label{fig:ctf_human_discussion_overview}
\end{figure*}

\section{CTF Game Design and Build} 
\label{sec:ctf_design_build}

We designed and built a CTF game in order to use it for our team assembly and the initial role assignments. The CTF game performances across different areas of functional role provided us with the information of the participants' strengths in those areas. The six role/skill categories were designed for the team project required in the class and are shown in Figure~\ref{fig:ctf_roles}. Capturing flags and earning the corresponding scores involved both knowledge/Internet search engine searching and hands-on software-based tasks, ranging from identifying a command to analyzing code or data. 

We built on the open-source CTFd platform because of its feature-rich design and easiness to develop and use~\cite{karagiannis_analysis_2020}. Our CTF game has also been alpha-tested by our team and beta-tested by four participants not involved in the research, development, and execution of this project. The participants played the CTF game for one hour (while the CTF is designed for 3-4 hour duration) on the first day of the Virtual Teams course. Due to the short duration, the CTF has been designed for a minimal learning curve from the participants and only required a computer, Internet connectivity/browser, a video-conferencing platform (Zoom for real-time interactions and coordination for the Coordination flags), and a camera (for QR code for a steganography problem). 

While it was an individual game (i.e., the participants had their own scores), the CTF was designed for better engagement and real-time coordination and interactions between the participants. Specifically, the Coordination category had flags involving real-time interactions among the participants (e.g., using their pre-shared codes to collectively solve a flag or finding a participant satisfying a condition). The CTF platform also showed a real-time scoreboard (the final scoreboard, excluding the identifiers/nicknames, is shown in Figure~\ref{fig:ctf_scores}) and a compressed version of the real-time scores and top performers in their main screen accessing and solving the challenges. In order to encourage trying and tackling multiple role/skill categories, the CTF unlocked more advanced flags if easier flags were solved and offered bonus incentive points for solving a greater number of role/skill categories (these incentive points have not been used in our algorithms for team assembly and initial role assignments in Section~\ref{sec:ctf_for_team_construction}). 
Our previous publication~\cite{zhang2022} describes in greater details about our design of the CTF to incorporate interaction and coordination on the CTF flags, which is a novel approach and design for the CTF games; this paper builds on our previous work but focuses on the CTF application for the team construction and facilitation after the CTF design/build, and the rest of the section does not overlap with our previous publication.

\section{CTF for Team Assembly and Initial Role Assignment}
\label{sec:ctf_for_team_construction}
Given that the CTF game can serve as an objective method for observing and measuring individuals' skills and expertise levels, we used the CTF performance data to assemble teams and assign the functional roles within each team. For the role assignment, the CTF provided a starting point, from which the team members could further discuss (e.g., share other inputs and relevant information such as preferences) and revise each member's roles and deliverables (Section~\ref{sec:human_discussion_role_assignment}). 

\subsection{Framework for Using CTF for Team Assembly and Role Assignment}
Below we present a mathematical framework for using the CTF for team assembly and role assignment. \emph{\textbf{Participants:}}
There is a set of $p$ number of participants $\mathcal{P}$. 
We use an integer index to identify the participants, as opposed to the participants' names/nicknames, for the participant anonymity, e.g., if there are fifteen participants ($p=15$), then $\mathcal{P} = \{1,2,...,15\}$.
\emph{\textbf{Roles:}} There is a set of $r$ functional roles $\mathcal{R}$. 
For example, in our CTF game, $\mathcal{R}$ is a subset of \{ IN, DE, AN, IM, TE, CO \}, 
since the CTF game implementation designed for engineers and cybersecurity students included six roles of Investigation (IN), Design (DE), Analysis (AN), Implementation (IM), Testing and Evaluation (TE), Coordination (CO) and the actual team construction and role assignment can require/cover less than those six roles. 
\emph{\textbf{Teams:}}
We form $n$ number of teams which form a set $\mathcal{N}$ and, from each team, we assign one functional role to each participant. 
\emph{\textbf{CTF Score (CTF Result):}}
The CTF objectively measures the participant scores for each functional role to indicate the skill and expertise level. The score matrix $S$ is of size $p \times r$, where the element on the $i$-th row and $j$-th column corresponds to the $i$-th participant (the $i$-th element in $\mathcal{P}$) and the $j$-th role (the $j$-th element in $\mathcal{R}$), i.e., the $i$-th participant's score for the $j$-th role category. 
\emph{\textbf{Capacity of Participant/Team}:} A participant's capacity is the sum of its scores across all role categories. For example, if a participant can assume all the roles and has the bandwidth to do so (in our implementation for our study, we assume that executing a role is significant and limit this to one role as described in Section~\ref{subsec:CTF_implementation_rules}), then the capacity directly affects the participant's contribution to the team. A team's capacity is the sum of the CTF scores for all role categories for all team members (a subset of the entire participants). 

\emph{\textbf{CTF for Team/Role Assignment:}} We take the score matrix and assemble the team and then assign the roles within each team, as described in Figure~\ref{fig:ctf_human_discussion_overview}. The team assembly results in a vector $\vec{s'}$ of length $p$ and whose element is an element in $\mathcal{N}$ 
indicating to which team the participant belongs. The role assignment takes the team assembly result $\vec{s'}$ and the score matrix $S$ to assign roles to each team members, resulting in a vector $\vec{s''}$ of length $p$ and whose element is an element in $\mathcal{R}$. 
$\vec{s'}$ including the elements of team information is the team assembly solution, while $\vec{s''}$ including the elements of role information is the role assignment solution. (In Section~\ref{sec:human_discussion_role_assignment}, $\vec{s''}$ gets updated to $\vec{s'''}$, both of which have the same format for the role assignment.)

\begin{table*}[th]
\begin{center}
\footnotesize{
\caption{Team Knowledge Sharing Assignment Work Sheet for Team Discussion-Based Role Assignment (as presented to the students but compressed and re-formatted to fit this manuscript)} 
\label{table:team_discussion_assign}
\vspace{-0.1in}
\begin{tabular}{ |p{1.3cm}|p{1.3cm}|p{1.3cm}|p{1.3cm}|p{2.4cm}|p{2.4cm}|p{2.0cm}|p{1.3cm}|p{1.3cm}| } 
 \hline
 Team\newline Member’s\newline Name 
 & \textcolor{blue}{\textbf{Functional Role Pre-Assigned Based on CTF}}
 & \textbf{Academic}\newline\newline Major/Minor \newline\newline Relevant coursework
 & \textbf{Work \newline Experience}\newline\newline
 Part- and full-time Internships
 & \textbf{Research \newline Abilities}\newline\newline
 Knowledge of current literature; Literature search; Research Integration; System analysis
 & \textbf{Software and \newline Hardware Skills}
 \scriptsize{\newline\newline Experience in programming languages, software tools, or hardware (e.g., networking, router or server, embedded platforms)}
 & \textbf{Communication Skills}\newline\newline
Leadership experience; Teamwork skills; Relationship management
 & \textbf{Other \newline Areas of\newline Expertise}
 & \textcolor{red}{\textbf{Final \newline Assigned Functional Role}}
 \\ 
 \hline
 1. &  &  & & & & & & \\ 
  \hline
 2. &  &  & & & & & & \\ 
  \hline
 3. &  &  & & & & & & \\ 
  \hline
 4. &  &  & & & & & & \\ 
 \hline
 5. &  &  & & & & & & \\ 
 \hline
\end{tabular}
}
\end{center}
\end{table*}

\subsection{CTF for Team/Role Assignment Implementation and Rules} 
\label{subsec:CTF_implementation_rules}

We implemented the CTF-informed team assembly and role assignment for our study. While we explain our implementation and envision it being directly applicable to many team scenarios, other implementations can adapt the framework in Section~\ref{sec:ctf_for_team_construction} in different ways (e.g., different metric and algorithms) to better accommodate their distinct team requirements and context. 

\noindent \textbf{Solution Feasibility} \hspace{0.15in}
The following are the set of assumptions and rules we used for our framework and thus helped define the solution feasibility. 
1) \textbf{Roles are essential.} Once the set or the subset of roles is established for each team (setting the expectations and requirements for a team), all roles need to be filled. Otherwise, the team does not meet the expectations, is incomplete, and not fully functional.  
2) \textbf{Executing the roles is substantial.} One person can fulfill and execute one role. Having one person execute multiple roles is excessive and infeasible. 
3) \textbf{Roles require expertise.} Roles require positive CTF score in the role. The participants not demonstrating any capability in the role and having zero point is not appropriate for the role.
4) \textbf{Participants can help each other.} Multiple people can work on the same role in a team.

These assumptions yield the feasibility requirements for the team assembly and role assignment, i.e., viable solutions satisfying the assumptions exist. For example, the feasibility requires that the number of teams be smaller than the number of total participants divided by the number of roles so that there are at least $r$ number of participants in each team to take on all the roles, i.e., $n \leq \left \lfloor \frac{p}{r} \right \rfloor$ where $\lfloor x \rfloor$ is the floor of $x$ or the greatest number smaller than $x$. Also, the number of people having the capability/expertise for any role needs to exceed the number of teams because, otherwise, there is at least one team which is incomplete and does not have the appropriate participants for the required roles. 

\noindent \textbf{Solution Selection} \hspace{0.15in}
In team assembly, we prioritized the fairness between the teams in their expertise and capabilities, i.e., reduce the variance across the teams, and reduced the expertise/capability discrepancy across the team members within the team. After assembling the team members, we assigned the roles to the team members to maximize the \emph{team score} aggregating the CTF scores for the member/role combinations while assuming zero scores for the roles not assigned. 

\noindent \textbf{Team Assembly} \hspace{0.15in}
Each team took turn in selecting the participant with the highest score in the remaining/non-selected participants. The turn taking reversed in direction in each round of selection, e.g., the team selecting the last in the previous round selects the first in the next round. 
We have considered an alternative scenario of assembling teams so that it minimizes the between-team variance in capacities, but such an approach increases the between-member variance in each team. The algorithm for our approach can also find a solution in polynomial time in O($p$). We further analyze our team assembly approach in Section~\ref{subsec:CTF_results}. 

\noindent \textbf{Initial Role Assignment} \hspace{0.15in}
Given each team resulting from the Team Assembly, 
we applied the Hungarian algorithm to assign the roles to the team members. Hungarian algorithm~\cite{hungarian_algorithm1955} is a popular combinatorial optimization algorithm for assignment and finds the optimal solution in polynomial time. Assuming that we keep only the score of a participant for the role that he/she gets assigned (e.g., the scores for the other roles for the participant effectively becomes zero), Hungarian algorithm in our implementation maximizes the aggregate score/contributions from the participants. Since the Hungarian Algorithm minimizes the cost, we pre-processed $S$ 
to generate another matrix, which is a sub-matrix of $S$ (selecting only the participants/rows within the team according to $\vec{s'}$), negating the scores for all elements (i.e., if each element is $x$ and $\alpha$ is the maximum score element in the newly generated matrix, we replace it with 
$\alpha-x$) and appending extra zero-element columns to make it a square matrix (if the number of team members exceed the number of roles).

\section{Team Discussion for Role Assignment}
\label{sec:human_discussion_role_assignment}

After assembling teams and assigning roles to each team member based on the CTF-based algorithm, we implemented a team-based discussion as part of the two-pronged approach to team knowledge sharing. Students as a team were instructed to submit in writing their responses to the following prompts: (1) reflect on their role pre-assigned based on the CTF and how accurately it represented their expertise, (2) share their backgrounds and expertise beyond what they have demonstrated in the CTF exercise by filling out the worksheet in Table~\ref{table:team_discussion_assign}, (3) consider if their initially assigned roles change based on the team discussion, and (4) reflect on how both the CTF and team discussion helped them recognize and utilize team members' knowledge for the project. In Section~\ref{subsec:results_qualitative} and Section~\ref{subsec:results_quantitative}, 
we report the results of a qualitative analysis of student teams' submission of this assignment as well as a quantitative comparison of course learning outcomes between the students who experienced the CTF exercise and those who did not.     


\section{Results and Analyses} 

\subsection{CTF Results} 
\label{subsec:CTF_results}

In this section, we describe the results for CTF session, team assembly, and initial role assignment for our beta-testing involving 4 participants (simpler) and the actual CTF session involving 10 participants (larger number and matrices). While we show the variable values for the simpler beta-testing session, for the later CTF session results, we omit the variable assignments and values for the space limitation (e.g., $S$ is a 10$\times$10 matrix), but rather focus on the description and insights of the results and analyses. 
While the beta-testing CTF session was solely for testing and fine-tuning the CTF itself, as described in Section~\ref{sec:ctf_design_build}, the actual CTF session with the participants in the Virtual Teams course resulted in team assembly and role assignments for all participants.  

\noindent \textbf{CTF Results for Beta-Testing (4 Participants, 4 Roles, and 1 Team)} \hspace{0.15in}
From the beta-testing session including four participants ($p=4$), we formed one team of four team members (all participants are in one team, i.e., $n=1$) where the team requires the roles of Investigation (IN), Design (DE), Implementation (IM), and Coordination (CO), i.e., we used $r=4$, $\mathcal{R}_1$ = \{ IN, DE, IM, CO \}. The selection of the roles is application and context dependent, for example, the Analysis (AN) and Testing and Evaluation (TE) roles can be outsourced or be added in a later stage for expanding the team. As a result of the CTF game, 
$S = \left [ \begin{smallmatrix} 23 & 257 & 83 & 256 \\ 103 & 60 & 20 & 290 \\ 10 & 150 & 61 & 238 \\ 50 & 141 & 61 & 0 \end{smallmatrix} \right ] $.
The team assembly resulted in all the four participants being in the same team (team ``1'' in this case), i.e., $\vec{s'} = [1,1,1,1]^T$. The initial role assignment to be $\vec{s''} = [\mbox{DE}, \mbox{IN}, \mbox{CO}, \mbox{IM}]^T$, i.e., the first participant takes on the Design role; the second participant takes on the Investigation role; the third participant takes on the Coordination role; and the fourth participant the Implementation role. The team score aggregating the scores for these participant/role combinations while assuming zero scores for the roles not assigned (in contrast to the ``team capacity'') is: 257+103+238+61=659. 

The required roles for the team affect the team assembly and the initial role assignment. 
For example, if we have greater flexibility in the roles needed for the team and can choose $\mathcal{R}_2$ = \{ DE, AN, IM, CO \} replacing IN with AN, then we actually can have greater team performance in the aggregate score. The CTF scores become: 
$S = \left [ \begin{smallmatrix} 257 & 97 & 83 & 256 \\ 60 & 0 & 20 & 290 \\ 150 & 10 & 61 & 238 \\ 141 & 55 & 61 & 0 \end{smallmatrix} \right ] $. The team assembly resulted in still the same team (team ``1'' in this case), i.e., $\vec{s'} = [1,1,1,1]^T$. The initial role assignment resulted in $\vec{s''} = [\mbox{DE}, \mbox{IM}, \mbox{CO}, \mbox{AN}]^T$. Compared with the earlier case with $\mathcal{R}_1$  
the aggregate team scores increases from before (659) to 257+290+61+55=663 using $\mathcal{R}_2$. The third participant also had higher role-aggregate score than the second participant, which would have been different if there were multiple teams. 

\noindent \textbf{CTF Results for Our Main Study (10 Participants, 5 Roles, and 2 Teams)} \hspace{0.15in}
From the CTF session at the start of the Virtual Teams course involving ten participants ($p=10$), we formed two teams of five team members each (i.e., $n=2$) where the team required the roles of Investigation (IN), Design (DE), Analysis (AN), Implementation (IM), and Coordination (CO), i.e., we used $r=5$, $\mathcal{R}$ = \{ IN, DE, AN, IM, CO \}. The selection of the roles is application and context dependent, for example, the Testing and Evaluation (TE) roles can be outsourced or be added later when expanding the team. As a result of the CTF game, the scores for each participant for each role is given and $S$ constructed (omitted for space reason). 
Two teams with feasible/valid solutions for team assembly and initial role assignment were constructed where, for each team, every team member took a distinct role. The team score aggregating the scores for these participant/role combinations was 179+350+115+81+264=989 for Team 1 and 253+325+95+116+8=797 for Team 2. 
This team assembly and the initial role assignment were used for the rest of the Virtual Teams Course team exercises, including the further role assignment for updating the roles. 

\begin{figure*}
	\centering
	\begin{subfigure}{.29\textwidth}
		\centering
		\includegraphics[width=1\textwidth]{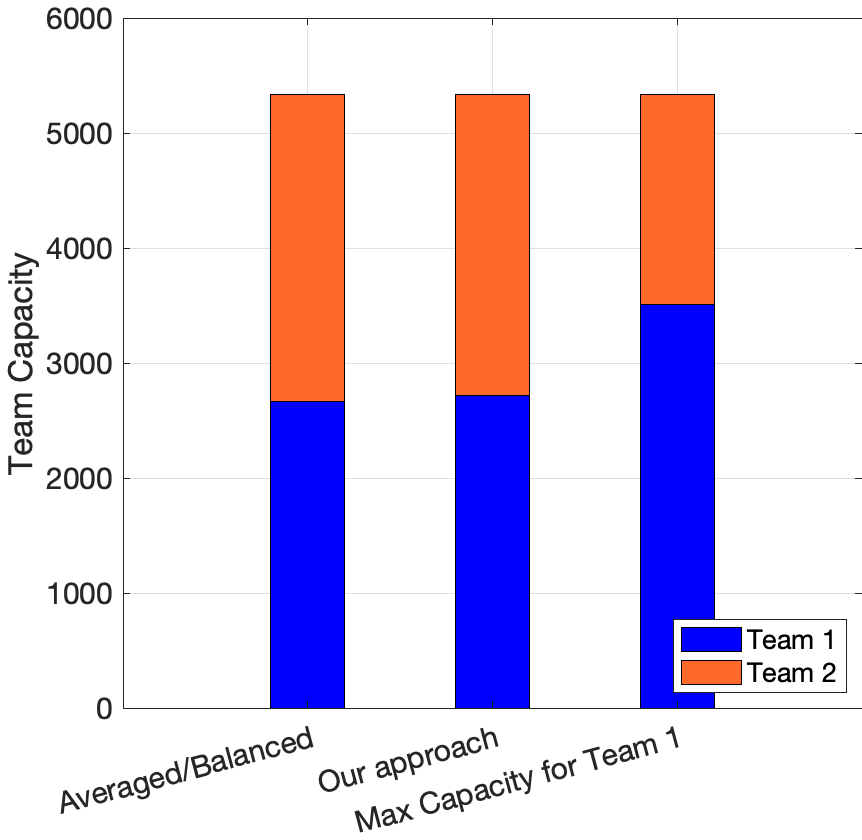}
		\caption{CTF results: team capacity after CTF-based team assembly}
		\label{fig:team_capacity}
	\end{subfigure}
	\quad
	\begin{subfigure}{.29\textwidth}
		\centering
		\includegraphics[width=\linewidth]{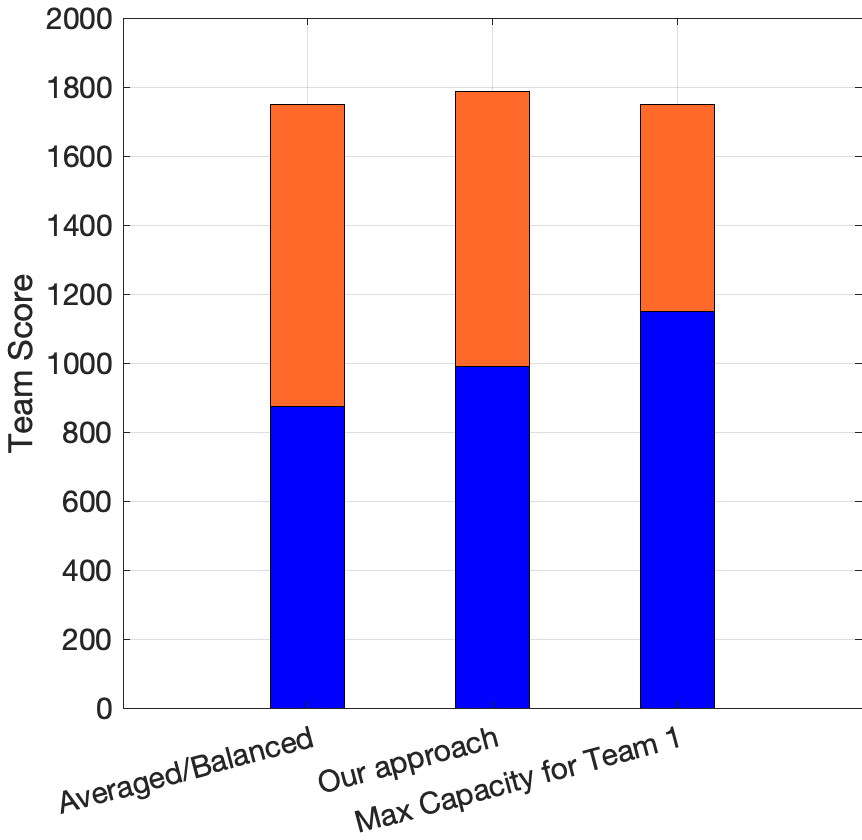}
		\caption{CTF results: team score after CTF-based role assignment}
		\label{fig:team_score}
	\end{subfigure}%
	\quad
	\begin{subfigure}{.36\textwidth}
		\centering
		\includegraphics[width=\linewidth]{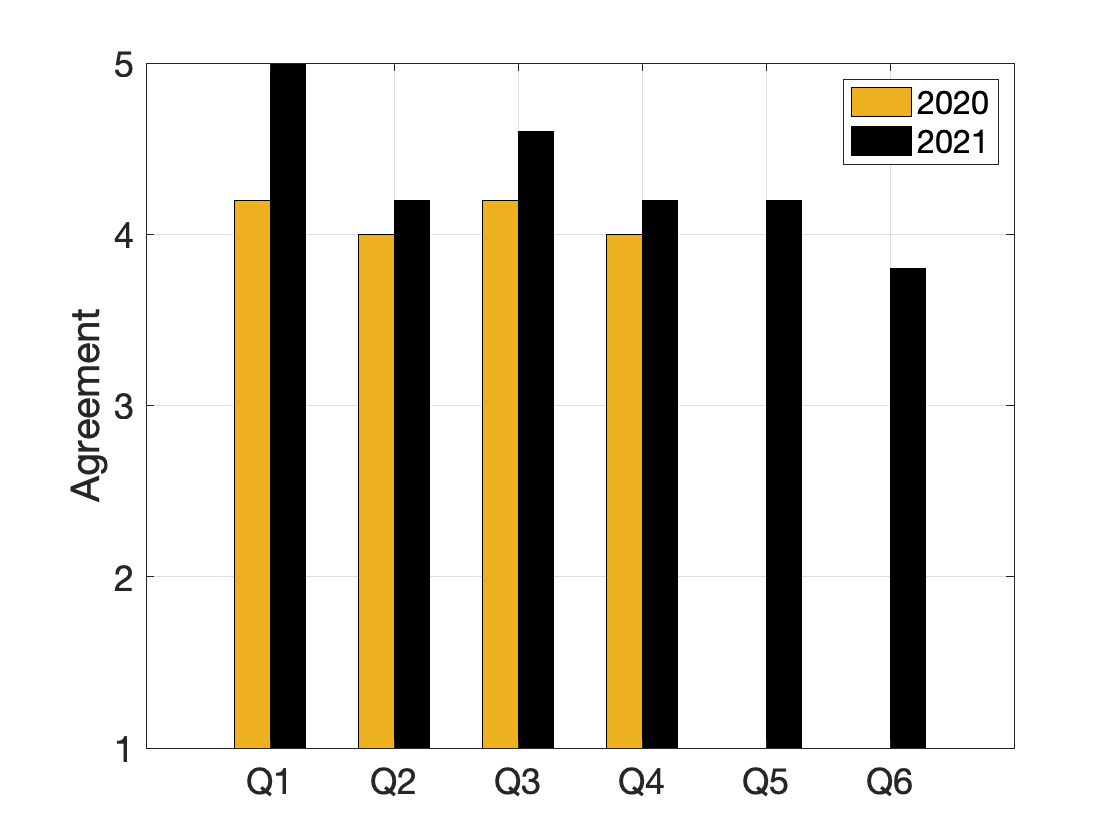}
		\caption{Evaluation results comparing 2020 (no CTF) vs. 2021 (with CTF-based team assembly and initial role assignment). Q5 and Q6 are specific to the CTF components newly added in 2021.} 
		\label{fig:survey_results}
	\end{subfigure}
	\caption{Results and Analyses} 
	\label{fig:test}
\end{figure*}

\subsection{CTF Algorithm Comparison Analysis}
\label{subsec:ctf_assembly_comparison}

We analyzed the CTF results in team assembly and initial role assignments. 
While the role assignment had a clear goal and definition for optimization to maximize the team score and to maximize the aggregate scores/contributions from the team members, the team assembly incorporated the trade-off between increasing the team capacity and decreasing the spread/discrepancies across the team members within a team in Section~\ref{subsec:CTF_implementation_rules}. Therefore, we compared our approach with the following alternatives: \emph{Averaged/Balanced Case)} tries all possible team-assembly cases and computes the average across those cases, while \emph{Max Capacity for Team 1} maximizes and prioritizes the capacity of Team 1 while the rest of the lower-capacity participants become Team 2. The Averaged/Balanced Case yielded the same capacities for both Team 1 and Team 2 and corresponded to the bound for making the team as fair and balanced as possible; however, this average case was infeasible as it includes capacities and scores in fraction. 
Figure~\ref{fig:team_capacity} shows the team capacity (the sum of all of the participant capacities) after the team assembly. It shows that our approach was close to the Averaged/Balanced Case where Team 1 only outperformed by $\frac{2721}{2665.5}-1 = 2.0821\%$ and Team 2 only lagged by $\frac{2610}{2665.5} = -2.0821\%$. In contrast, the most imbalanced case for Max Capacity for Team 1 had a discrepancy of +31.645\% for Team 1 and -31.645\% for Team 2.

We also verified that, once the teams are assembled, the polynomial-time-efficient Hungarian algorithm maximizes the team score by comparing it with all other possible cases for role assignments (exhaustive search on the role assignments). Figure~\ref{fig:team_score} shows the team scores (only counting and summing the participant scores for his/her role) after the Hungarian Algorithm was applied for role assignments. While the Averaged/Balanced Case had the team score difference of $\frac{875.18}{875.18}-1 = 0\%$ between Team 1 and Team 2, our approach had a difference of $\frac{989}{797}-1 = 24.090\%$ between Team 1 and Team 2 team scores and Max Capacity for Team 1 has a difference of $\frac{1150}{598}-1 = 92.308 \%$ between the two teams' scores.

\subsection{Qualitative Analysis of Team Discussions} 
\label{subsec:results_qualitative}

In their qualitative reflections on the CTF exercise and the team discussion for the course assignment, both teams indicated that the initial role assignment based on the CTF performance was an accurate representation of their team members' expertise. One team made no adjustment to their initial role assignment, while the other team made only one role swap between two members. Students' reflections suggest that the team discussion facilitated expertise recognition and role assignment in two ways. Firstly, the team discussion helped them confirm and reinforce each member's expertise demonstrated in the CTF exercise. For instance, one team stated that "we discussed our backgrounds in-depth and learned that our backgrounds lent themselves to the functions we had been assigned by the algorithm." The other team also stated that "the CTF provided a starting point to compare our roles against our skills. The team discussion was an effective mechanism for hashing out any other details." Secondly, the team discussion helped teams fine-tune their understanding about the members' relative expertise in different role categories and offered an opportunity for members to express their preferences. The team who switched roles between two members stated that "our discussion helped us understand where each person would fit best when also considering personal preference along with skill set. We also discussed where we felt the team had holes or less experience and where we had overlapping experience. [Member A]’s experience in classwork with software projects is a better fit for Implementation. [Member B]’s work experience as an information security analyst is better suited for Analysis." These student reports suggest that the algorithmic and communicative processes are complementary and in combination enhance the accuracy of expertise recognition and role assignment.

\subsection{Quantitative Analysis of The Impact of CTF on Team Collaboration} 
\label{subsec:results_quantitative}

To gauge the impact of the CTF exercise on students' learning in team collaboration skills, we compared student responses to an end-of-the-course questionnaire between the course with and without the CTF component. All other elements of the two versions of the course (e.g., reading, individual assignments, lectures, discussion forums) were exactly the same, and the incorporation of the CTF exercise was the only differentiating factor. In the questionnaire, students were asked to indicate their level of agreement in a scale of 1-5, 1 being "strongly disagree" and 5 being "strongly agree", to four statements: Q1 stated "My understanding of team processes has improved after completing this course.", Q2 "After taking this course, I feel more confident about my own team work skills.", Q3 "After taking this course, I recognize the value of team efforts in the field of cybersecurity more significantly.", Q4 "The course contents (e.g., readings, lecture, assignments) were applicable to my work." Two additional questions specific to the CTF exercise were asked to the students who experienced the CTF exercise: Q5 stated "The results of the CTF exercise helped me learn about my team members' backgrounds and expertise.", Q6 "The results of the CTF exercise helped my team members assign functional roles to each other effectively in the team assignment."

The comparison\footnote{One-way ANOVA was performed to test the statistical significance when comparing means between the two groups. However, due to the small sample size (N=5 in each group), the statistical procedure is deemed inappropriate, and therefore, its results are not fully reported in the paper. The results of one-way ANOVA are available upon request.} of the means between the two groups shows a positive impact of the CTF exercise in all four questions Q1-Q4, as shown in Figure~\ref{fig:survey_results}. Students who experienced the CTF exercise showed higher ratings in their understanding of team processes (mean with CTF=5.0, mean without CTF=4.2), their confidence in team collaboration skills (mean with CTF=4.2, mean without CTF=4.0), the appreciation for team efforts (mean with CTF=4.6, mean without CTF=4.2), and the applicability of course materials (mean with CTF=4.2, mean without CTF=4.0) than those who did not experience the CTF exercise. Students' average ratings for the CTF exercise (Q5 mean=4.2; Q6 mean=3.8) also indicate a positive impact of the CTF exercise on expertise recognition and role assignment. 


\section{Conclusion}
\label{sec:conclusion}


Team collaboration skills are one of the most desired competencies in the contemporary workforce including the field of cybersecurity. While the benefits of teamwork lie in the diversity of skill sets that individual members can bring to collective tasks, teams often fail to recognize individual members' strengths and expertise or to fully utilize them. In an effort to enhance expertise recognition whereby team members can assign roles to each other in line with their areas of expertise and share each other's expertise more effectively, we examined in our work if the two-pronged approach---a CTF-based algorithm and a team discussion---positively affects expertise recognition in teams. Our results show that our computer-human process is effective for team assembly and role assignment and made a positive impact on the participants' learning of team collaboration skills and the course materials. 

\section*{Acknowledgment}
This material is based upon work supported by the National Science Foundation under Grant No. 1922410. 

\bibliographystyle{ACM-Reference-Format}
\bibliography{ctf_team_role}

\end{document}